\documentclass[aps,prd,preprint,superscriptaddress,nofootinbib,showpacs]{revtex4}
\usepackage{graphicx}
\usepackage{amssymb}
\usepackage[centertags]{amsmath}
\usepackage{ulem}
\usepackage{color}

\newcommand{\beq}{\begin{equation}}
\newcommand{\eeq}{\end{equation}}
\newcommand{\wt}{\widetilde}
\def\bea{\begin{eqnarray}}
\def\eea{\end{eqnarray}}

\def\ld{\lambda}

\begin{document}

\title{ Oscillating Asymmetric Sneutrino Dark Matter\\ from the 
Maximally  $U(1)_L$ Supersymmetric Inverse Seesaw}

\author{Shao-Long Chen}
\email{chensl@mail.ccnu.edu.cn}
\affiliation{Key Laboratory of Quark and Lepton Physics (MoE) and Institute of Particle Physics,
Central China Normal University, Wuhan 430079, China }
\affiliation{Center for High Energy Physics, Peking University, Beijing 100871, China}

\author{Zhaofeng Kang}
\email{zhaofengkang@gmail.com}
\affiliation{ School of Physics, Korea Institute for Advanced Study, Seoul 130-722, Korea}

\begin{abstract}

The inverse seesaw mechanism provides an attractive approach to generate small neutrino mass, which origins from a tiny $U(1)_L$ breaking. In this paper, we work in the supersymmetric version of this mechanism, where the singlet-like sneutrino could be an asymmetric dark matter (ADM) candidate in the maximally $U(1)_{L}$ symmetric limit. However, even a tiny $\delta m$, the mass splitting between sneutrino and anti-sneutrino as a result of the tiny $U(1)_{L}$ breaking effect, could lead to fast oscillation between sneutrino and anti-sneutrino and thus spoils the ADM scenario. We study the evolution of this oscillation and find that a weak scale sneutrino, which tolerates a relatively larger $\delta m\sim 10^{-5}$ eV, is strongly favored. We also investigate possible natural ways to realize that small $\delta m$ in the model.

\end{abstract}
\pacs{12.60.Jv, 14.70.Pw, 95.35.+d}

\maketitle

\section{Introduction}
It is well established that dark matter (DM) accounts for about one quarter of the total energy in the Universe. The nature and origin of DM still remain unclear. The popular dark matter candidates are used to be characterized as weakly interacting massive particles (WIMPs). The WIMPs are CP-symmetric and freeze out of the thermal equilibrium when the Universe cools down, naturally providing the correct relic density of dark matter, called the ``WIMP miracle''.  Alternatively,  asymmetric dark matter (ADM) provides another way to understand the DM puzzle and draw much attention~\cite{Barr:1990ca, Petraki:2013wwa, Zurek:2013wia,Buckley:2010ui} 
(or its variant metastable asymmetric particle~\cite{Kang:2011ny, Barr:2015lya, Unwin:2012rp}). Similar to the baryons, the asymmetric DM abundance is fixed by the dark matter's charge asymmetry,  with a conserved 
symmetry $U(1)_{\rm DM}$ acting as DM number. Moreover, the asymmetries in dark sector and baryon sector might be dynamically connected, which supplies a natural way to explain the coincidence of the baryon and dark matter densities.  
 
The smallness of neutrino mass is another puzzle which drives us to go beyond the standard model. It is tempting to build a 
bridge between the dark matter, particularly the ADM and neutrino physics, which is closely associated to 
the lepton number $U(1)_L$. In the canonical seesaw mechanism, which provides a natural way to generate 
tiny neutrino masses, large lepton number $U(1)_L$ breaking effects are provided
by the heavy Majorana mass terms of the right-handed neutrinos. By contrast, the inverse seesaw 
mechanism~\cite{Mohapatra:1986bd} attributes the smallness of neutrino mass to a tiny $U(1)_L$ breaking term, 
potentially allowing for a highly conserved $U(1)_L$.

In a maximally $U(1)_L$ supersymmetric inverse seesaw standard model (M$L$SIS), an ADM candidate, 
the sneutrino being the lightest sparticle, is nicely presented~\cite{Kang:2011wb} (see another example~\cite{Mitropoulos:2013fla}). The model distinguishes from several relevant studies in literature, e.g., a study of real rather than complex sneutrino DM without maximal $U(1)_L$~\cite{Arina:2008bb}; a complex but not asymmetric sneutrino DM~\cite{An:2011uq, Guo:2013sna, Choi:2013fva}; a seemingly asymmetric but actually symmetric sneutrino, after taking into account the effects like neutralino-mediated washing-out and DM-anti-DM oscillating which were missed before~\cite{Hooper:2004dc, Page:2013fva}.

In this work, we point out that a remarkable feature of the sneutrino ADM provided in the M$L$SIS model is that 
DM and antiDM are oscillating~\cite{oscillating1,Tulin:2012re, oscillating} (such phenomena was mentioned 
before in a few papers~\cite{Kang:2011wb,osci0}), as a consequence of tiny $U(1)_L$ breaking. 
We study the evolution of the sneutrino asymmetric DM in detail.  We find that the sneutrino ADM 
is strongly favored to be around the weak scale instead of the GeV scale like in most ADM models. 
In addition that, we notice that the ADM will evolve into chemical equilibrium with 
neutrino via DM charge violating scattering process mediated by neutralinos, which could wash out the asymmetry 
during ADM freeze-out.  To avoid it, the ADM should be sufficiently singlet-like. 

This work is organized as follows. In Sec. II we present a dynamical model with supersymmetric theory
for the inverse seesaw mechanism. In Sec. III, we study the evolution processes of the ADM in detail.
We give numerical results to illustrate the oscillating effects. We conclude in Sec. IV.


 \section{Maximal $U(1)_L$ supersymmetric inverse seesaw (M$L$SIS)}\label{model}
The inverse seesaw mechanism~\cite{Mohapatra:1986bd} provides an elegant way to understand the smallness of neutrino mass. 
The tiny Majorana masses of the active neutrinos which break lepton number by two units are 
consequences of slight $U(1)_L$ breaking. The minimal implementation is introducing 
a pair of pseudo-Dirac particle $(N, N^c)$ with a tiny Majorna mass term $M_{N}$, which breaking the 
$U(1)_L$ explicitly. In the supersymmetric version, the superpotential is given by
\begin{align}\label{superpotential}
W=y_NH_uLN^c+{m_N}N^cN+\frac{M_{N}}{2}N^2.
\end{align}
The superfields components are denoted as $N^c=(\wt \nu_R^*,\nu_R^\dagger)$ and $N=(\wt \nu_L^{\prime},\nu_L^{\prime})$, where
$\nu_L^{\prime}$ and $\nu_R$ carry lepton number $+1$ as $\nu_{L}$. 
To illustrate our main idea, we consider one family of neutrino for simplicity and implications 
of multi-family will be commented if necessary. The corresponding soft terms are
\begin{align}\label{softer}
-{\cal L}^{soft}=&\left( m_{\wt L}|\wt L|^2+m_{\wt \nu_L^{\prime}}|\wt \nu_L^{\prime}|^2+m_{\wt \nu_R}|\wt \nu_R|^2\right)\cr
&+y_NA_NH_u\wt L\wt \nu_R^*+B_m{m_N}\wt \nu_L^{\prime}\wt \nu_R^*+\frac{B_{M}M_{N}}{2}(\wt \nu_L^{\prime})^2 + h.c.,
\end{align}
where the soft SUSY-breaking parameters $A_N$, $B_m$, etc., are assumed to be real and around the weak scale. 

In the flavor basis $(\nu_L,\nu_R^\dagger,\nu_L^{\prime})$, the neutrino mass matrix is given by
\begin{align}\label{}
{\mathcal M}_{\nu}=\left(\begin{array}{ccc}
              0 & m_D & 0 \\
              m_D & 0 & m_N \\
              0 & m_N & M_{N}
            \end{array}\right),
\end{align}
with the Dirac neutrino mass $m_D=Y_N\langle H_u^0\rangle$. 
In the case when $m_N/m_D\gg 1$, the lightest mass eigenvalue is given by
\begin{align}\label{light}
m_{\nu}^{eff}=-\frac{m_D^2}{m_N^2+m_D^2}M_{N},
\end{align}
which, as expected, is proportional to the $U(1)_L$-breaking Majorana mass term. 
The lightest neutrino is dominated by the active neutrino
and contain a small fraction of $\nu_L'$, 
\begin{align}
\nu_{1}\approx \cos{\theta_\nu} \nu_L- \sin{\theta_\nu} \nu_L^{\prime},
\end{align}
with $\sin{\theta_\nu}\approx m_{D}/m_{N}\ll 1$. The mixing $\theta_{\nu}$ between $\nu_{L}$ and $\nu_{L}^{\prime}$ 
will introduce non-unitarity effects which may be observable in future experiments~\cite{Antusch:2014woa}. 
To avoid too large non-unitarity effects we set the mixing ${\theta_\nu}\sim m_{D}/m_{N}\lesssim {\mathcal O}(10^{-2})$. 
The light neutrino mass is naturally small due to this suppressing factor and the smallness of $M_{N}$, which is dynamically 
generated in the model, maintaining $U(1)_L$ to the most extent.

Note that the mixing between $\nu_{L}$ the $\nu_R^\dagger$ is negligible since it is severely suppressed by 
$M_Nm_D/m_N^2\sim m_{\nu}^{eff}/m_{D}$. The remaining two mass eigenstates are  $\nu_{2,3} \approx
\frac{1}{\sqrt{2}}\left( \pm  \nu_R^\dagger+ {\sin{\theta_\nu}}\nu'_L+\cos{\theta_\nu} \nu_L\right)$. They have 
almost degenerate masses $|M_{2,3}|=\sqrt{m_N^2+m_D^2}+{\cal O}(M_{N})\approx m_N$ and form a pair of
pseudo-Dirac fermions.  

\section{Oscillating asymmetric sneutrino dark matter}
In this section we study the interesting phenomenologies of the asymmetric sneutrino dark matter. 
First, we investigate the asymmetric sneutrino dark matter in the limit of exact $U(1)_L$. Then we turn on the tiny
$U(1)_L$ breaking term and study the DM-antiDM oscillation. 

\subsection{Asymmetric sneutrino dark matter in the $U(1)_L$ limit}
We choose the sneutrino instead of conventional neutralino to be the LSP dark matter candidate. 
Neglecting the tiny $U(1)_L$ breaking term,  it carries  lepton number and/or dark matter number, 
thus can be asymmetric DM.  

In the basis $(\wt \nu_{L},\wt \nu_R,\wt \nu_L')$, from Eq.~(\ref{superpotential}) and  Eq.~(\ref{softer}) the sneutrino mass squared matrix is given by
\begin{align}\label{mL2}
            m_{\wt \nu}^2\approx&\left(\begin{array}{ccc}
              m_{\wt L}^2+\frac{1}{2}M_Z^2\cos2\beta+m_D^2 &
\left( -m_D A_N+\mu m_D\cot\beta\right) & -m_D m_N \\
                & m_{\wt \nu_R}^2+m_N^2+m_D^2 & B_mm_N \\
               &  & m_{\wt\nu_L'}^2+m_N^2 
            \end{array}\right),
\end{align}
where the $\mu-$term is from $\mu H_uH_d$, which is not explicitly included in Eq.~(\ref{superpotential}). We also include the $D-$term contribution to the left-handed sneutrino. For simplicity, we assume all parameters to be real. Later we will see that the left-handed sneutrino is forced to almost decouple from other two sneutrinos, and hence we can make the good approximation
\begin{align}\label{rotation}
\wt \nu_L'\approx-\sin\wt\theta \wt \nu_1+ \cos\wt\theta \wt \nu_2,\quad \wt \nu_R\approx\cos\wt\theta \wt \nu_1+ \sin\wt\theta \wt \nu_2,
\end{align}
with $\wt\theta$ the mixing angle between the two singlet sneutrinos. 

The asymmetric DM scenarios provide very attractive ways to understand the coincidence between the relic densities of the dark 
and baryonic matters, $\Omega_{\rm DM}:\Omega_{b}\simeq 5:1$~\cite{Barr:1990ca}. In this work we assume that the matter-
antimatter asymmetry is generated through certain mechanism in the visible sector and then transferred into the dark sector.  
Chemical equilibrium dynamically connects the chemical potential $\mu$ for various particles and typically 
we get $\mu_{\rm baryon}\sim \mu_{\rm DM}$. At temperature $T$,  the asymmetry of particle $\phi$ with mass $m_\phi$ in the thermal bath
 can be expressed in terms of $\mu_\phi$ (the lower index will be ignored)~\cite{TEU}
\begin{align}\label{asymmetry}
n_+-n_-&=g\frac{T^3}{\pi^2}\frac{\mu_\phi}{T}\int_0^\infty dx
\frac{x^2\exp[-\sqrt{x^2+(m/T)^2}]}{\left(\theta+\exp[-\sqrt{x^2+(m/T)^2}]\right)^2} \nonumber\\
&= \left\{ \begin{array}{l}
f_{b}(m_\phi/T)\times \frac{gT^3}{3}\left(\frac{\mu_\phi}{T}\right),\quad {\rm(for\,\, bosons)} \\
f_{f}(m_\phi/T)\times \frac{gT^3}{6}\left(\frac{\mu_\phi}{T}\right),\quad {\rm (for\,\,fermions)}
\end{array}\right.
\end{align}
with $\theta=\pm1$ for fermion/boson. The Boltzmann suppression factor $f_{b,f}(m_\phi/T)$ denotes the threshold effect for heavy particle in the plasma.  It tends to 1 for particles in the ultra-relativistic limit $m_\phi\ll T$.

The key point is that the symmetric parts of both baryonic matter and DM will annihilate away and only the asymmetric parts 
survive. As a consequence, their number densities are connected. In this way the coincidence puzzle can be understood, 
given a proper DM mass. 
To see this, we consider the limit that the chemical equilibrium between two sectors breaks at $T_d$ which is much higher than DM mass (hereafter we define $x\equiv m_{\rm DM}/T$), and thus $f_{\rm DM}(x_d)\sim f_{\rm DM}(0)\approx 1$. Then from
\begin{eqnarray}\label{puzzle}
\frac{\Omega_bh^2}{\Omega_{\rm DM}h^2}=\frac{m_n}{m_{\rm DM}}
\frac{\mu_b}{g_{\rm DM}f_{\rm DM}(0)\mu_{\rm DM}}\approx \frac{1}{5}
\end{eqnarray}
with $m_n\simeq1$ GeV the nucleon mass, the DM is expected to be light, with mass around 5 GeV. In contrast, for the case of
 $x_d\gg1$, the residual ADM asymmetry will be suppressed by a factor $f(x_d)\ll1$ and the resulting DM mass will scale as 
 $5f^{-1}(x_d)$GeV, easily entering the TeV region~\cite{Buckley:2010ui}. We will see later in our model the sneutrino ADM is 
 favored to be in the heavy region. 

\subsection{Evolution of sneutrino ADM}
In this subsection we trace the evolution of sneutrino asymmetry.  As we assumed, 
the lepton number asymmetry has been generated at some high temperature 
and part of it has been transferred to the right-handed neutrino (RHN) $(N, N^{c})$ through the Yukawa interaction $y_NLH_uN$. 
There are several critical temperatures during the evolution of sneutrino asymmetry: (I) the out-of-equilibrium temperature of 
electroweak sphaleron process $T_{sph}$,  below which the connection between 
the lepton and baryon chemical potentials gets lost; (II) the chemical equilibrium decoupling temperature $T_d$
between dark matter and the visible sector, more concretely,  the leptons; (III) the dark matter freeze-out temperature $T_f\sim m_{\wt \nu_1}/20$. 

\subsubsection{$T\sim T_{sph}$: baryon number freeze-out}
Above the temperature $T_{sph}$ all Yukawa interactions are supposed to be in chemical equilibrium 
and thus three families of fermions share the same chemical potential.~\footnote{Charged lepton flavors in the SM or  models with minimal flavor violation are individually conserved, 
as is different to the quark sector where charged currents can drive different flavors share the same chemical potential. 
But in the presence of RHNs with appreciable family-interchanging 
Yukawa couplings $(y_{N})_{ij}$, leptonic chemical potential is putative common.} Moreover, we assume that $T_{sph}$ is lower than the EW phase transition critical temperature $T_c$, so the Higgs condensations lead that the Higgs neutral components have zero chemical potential. As a consequence, the left- and right-handed fermions (including RHNs), develop same chemical potential via the Yukawa interactions:
\begin{align}\label{}
\mu_{u_L}&=\mu_{u_R},\quad \mu_{d_L}=\mu_{d_R}\cr
 \mu_{e_L}&=\mu_{e_R},\quad \mu_{\nu_L}=\mu_{\nu_R}=\mu_{\nu_L^{\prime}}.
\end{align}
The $W$-boson mediated gauge interactions force the down and up components of the $SU(2)_L$ doublets to acquire chemical potential related as
\begin{align}\label{}
\mu(I_3=-1/2)=\mu(I_3=1/2)+\mu_W.
\end{align}
The electroweak sphaleron process is effective and thus the left-handed quarks and leptons are further forced to satisfy the relation:
\begin{align}\label{sp}
\mu_{u_L}+2\mu_{d_L}+\mu_{\nu_L}=3\mu_{u_L}+\mu_{\nu_L}+2\mu_W=0.
\end{align}
Eventually, the plasma should be QED neutral,
\begin{align}\label{QED}
3\mu_{u_L}-3\mu_{\nu_L}-9\mu_W=0.
\end{align}

For simplicity we take all sparticles, except for $k$ singlet-like sneutrinos, to be highly Boltzmann suppressed 
and thus do not contribute to total charge asymmetry. This approximation is reasonable viewing from the current null results 
of LHC searches for superparticles. The sneutrinos are brought into equilibrium with the thermal bath via the $y_N-$terms 
as well as the relevant soft terms. They have identical chemical potential with RHNs,
\begin{align}\label{}
\mu_{\wt \nu_R}=\mu_{\wt \nu_L'}=\mu_{\nu_L}.
\end{align}
Combining Eq.~(\ref{sp}) with Eq.~(\ref{QED}), all chemical potentials can be expressed in terms of the single variable $\mu_{u_L}$,
\begin{align}\label{}
\mu_{\nu_L}=-11\mu_{u_L},\quad \mu_{W}=4\mu_{u_L}.
\end{align}
Therefore we obtain the total baryon and ADM (sneutrino) number:
\begin{align}\label{}
B(T_{sph}) =& \frac{T_{sph}^2}{6}\left[ 2 \times 3 (2\mu_{u_L}+\mu_W)\right]=6T_{sph}^2\mu_{u_L},\cr
S_{\rm ADM}(T_{sph})=&\frac{T_{sph}^2}{3} k \mu_{\nu_L}=-\frac{11k}{3}T_{sph}^2 \mu_{u_L}.
\end{align}
Hereafter, baryon number will preserve the initial value $\eta_b\equiv {B}/{s(T_{sph})} =1.02\times10^{-10}$~\cite{baryon:as} with entropy density
\begin{align}\label{}
s(x)=&\frac{2\pi^2}{45}g_{*S} m_{\rm DM}^3x^{-3}.
\end{align}
$g_{*S}\approx g_{*}$ is the effective relativistic degrees of freedom, approximated to be $x-$independent.

\subsubsection{$T\sim T_d$: Washing-out effect}

Below $T_{sph}$ the quark and lepton sectors lose chemical equilibrium, but the dark sector (the RHN-like sneutrino sector) will keep chemical equilibrium with leptons until $T_d$. Below $T_{d}$ the asymmetry transferring between these two sectors ceases and DM number becomes separately conserved. In the M$L$SIS-like models, as pointed out by Ref.~\cite{Kang:2011ny}, there may exist DM charge violating scattering (CVS) processes mediated by neutralinos which could stay active even around the DM freeze-out temperature $T_f$. In other words,  for $T_d\lesssim T_f$, the CVS processes maintain chemical equilibrium between ADM and the active neutrinos, and consequently the ADM asymmetry is washed-out during freeze-out,~\footnote{Ref.~\cite{Ellwanger:2012yg} noticed a different way to wash-out asymmetry, DM charge violating annihilation 
like $\wt\nu_1\wt\nu_1\rightarrow \nu\nu$, which is also mediated by neutralinos. However, unlike the scattering process here, that process does not regenerate symmetric DM; it merely reduces the amount of the initial asymmetry.} i.e., the asymmetric component can contribute to the total DM relic density at most a subdominant fraction (See also other scenarios for this kind of phenomena~\cite{Boucenna:2015haa}.). 

The underlying reason is that ADM is not the lightest particle carrying lepton number/charge, so it
cannot retain its asymmetry unless the CVS processes are sufficiently suppressed. This is not trivial since the scattering happens 
between the non-relativistic DM and relativistic neutrino, whereas freeze-out is with respect to the annihilation of 
two non-relativistic DM, which is Boltzmann suppressed. The dominate CVS process is $\wt\nu_1\nu_1\leftrightarrow \wt\nu_1^*\bar\nu_1$, originating from the following effective Lagrangian: 
\begin{align}\label{}
-{\cal L}_{wash}=\frac{1}{2}M_i^2\bar \chi_i\chi_i+\left( y_{i1}\wt \nu_1^*\bar \chi_i P_L\nu_1+h.c.\right).
\end{align}
with $\chi_i$ denoting four Majorana neutralinos, which are related to the states in the interacting eigenstates via $\chi_i=Z^T_{ij}\psi_j$ with $\psi=(\wt B,\wt W^3,\wt H_d^0,\wt H_u^0)^T$.  The couplings $y_{i1}$ receive several contributions and here we ignore the parts involving gauge interactions, which is justified because we decouple the left-handed sneutrino $\wt \nu_L$. Then we have 
\begin{align}\label{}
y_{i1}\approx y_N \sin{\theta_\nu}\cos\wt\theta Z_{4i},
\end{align} 
where we have used the fact that the light neutrino $\nu_1$ is mostly left-handed. 
We will see that this feature leads to an extra kinematic suppression factor in the scattering rate.

\begin{figure}[htb]
\begin{center}
\includegraphics[width=5.2in,natwidth=10,natheight=15]{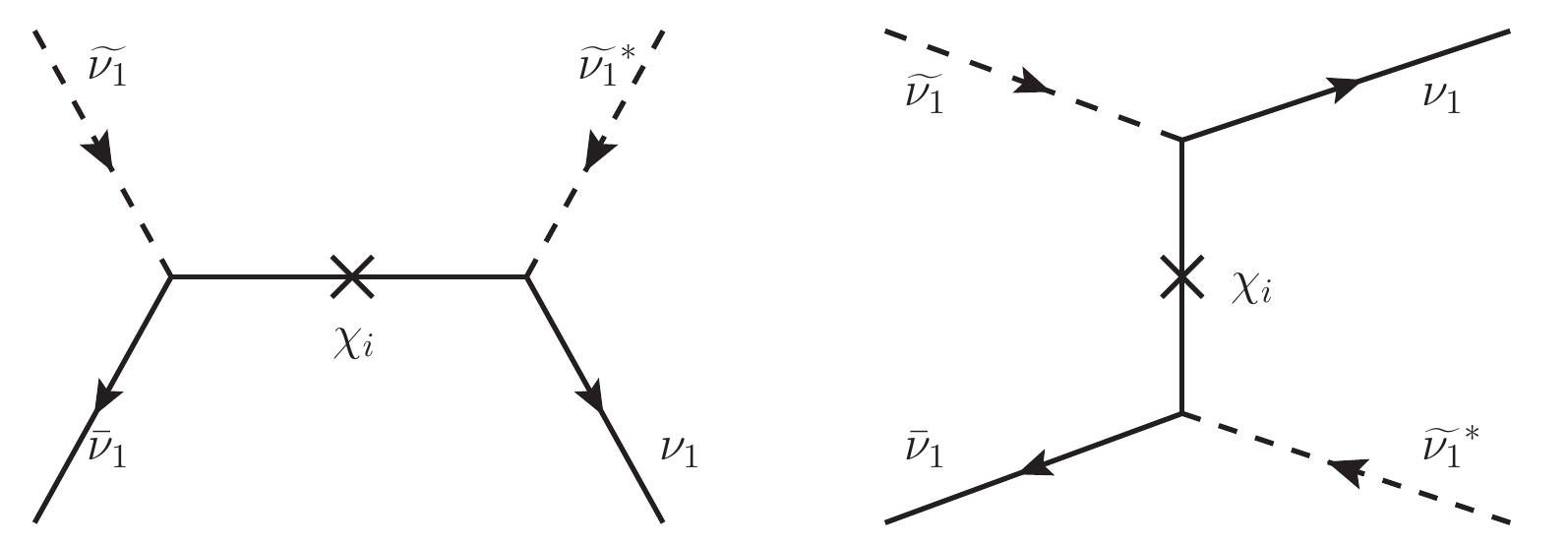}
\end{center}
\caption{{Neutralino-mediated  sneutrino dark matter charge violating process in the $s$- and $t$-channel.}}
\label{CVS}
\end{figure}

To save the sneutrino ADM from washing out, $y_{i1}$ should be small enough to 
ensure the CVS processes cease above $T_f$. 
There are $s-$ and $t-$channel contributions to the CVS process $\wt \nu_1\bar\nu_1\rightarrow \wt \nu_1^* \nu_1$, as shown in Fig.~\ref{CVS}. 
The squared amplitude is 
\begin{align}\label{}
|{\cal M}_s+{\cal M}_t|^2=2 |y_{i1}^2|^2k_2\cdot p_2 M_i^2\left(\frac{1}{s-M_i^2}+\frac{1}{t-M_i^2}\right)^2.
\end{align}
In the CM frame, for kinematics specified by the scattering between relativistic and non-relativistic particles, we have the following expressions:
\begin{align}\label{}
&k_2\cdot p_2\approx w_\nu m_{\wt \nu_1} \cos\theta,\quad s\approx m_{\wt \nu_1}^2,\cr
&  t=(p_2-k_1)^2\approx s+2m_{\wt \nu_1}w_\nu\cos\theta.
\end{align}
The total scattering cross section is given by
\begin{align}\label{}
\sigma_{\rm CVS}\approx \frac{ |y_{i1}^2|^2}{6\pi} \frac{w_\nu^4}{M_i^4 m_{\wt \nu_1}^2}.
\end{align}
Heavy neutralinos are favored to suppress the cross section.

The thermal scattering rate can be estimated in a way as used for the neutron decoupling from the heat bath 
via the scattering $n+\nu\leftrightarrow p+e^+$~\cite{TEU}. We obtain
\begin{align}\label{}
\Gamma_{\rm CVS}=\int_0^\infty dw_\nu  \sigma_{\rm CVS} (f_\nu v_\nu  g_{w_\nu})\frac{1}{1+e^{-w_\nu/T}},
\end{align}
with the phase volume element $ g_{w_\nu} dw_\nu=g_\nu w_\nu^2 /2\pi^2 dw_\nu$  ($g_\nu=2$ the internal degree of freedom of left-handed neutrino). $f_\nu={1}/({1+e^{w_\nu/T})}$ is the Fermi-Dirac statistics and $v_\nu=1$ is the velocity of neutrino. The Pauli exclusion effect in the thermal bath is taken into account by the factor ${1}/(1-e^{w_\nu/T})$. After integration we obtain
\begin{align}\label{}
\Gamma_{\rm CVS}=\frac{19845 \,{\rm Zeta}[7]}{4}\frac{ |y_{i1}^2|^2}{12\pi^3}\left(\frac{T}{M_i}\right)^4\left(\frac{T}{m_{\wt \nu_1}}\right)^2 T.
\end{align}
The numerical prefactor is 5002.7. The condition for CVS decoupling at $T_d$ is fulfilled 
as long as $\Gamma_{\rm CVS}(T_d)<H(T_d)\approx 5.5T_d^2/M_{\rm Pl}$, which set an upper bound on the couplings
\begin{align}\label{CVSbound}
{ |y_{i1}^2|^2}\lesssim0.41x_d \left(\frac{M_i}{T_d}\right)^4 \frac{m_{\wt \nu_1}}{M_{\rm Pl}}=0.33\times10^{-8}\left( \frac{M_i/m_{\wt \nu_1}}{10}\right)^4\left(\frac{x_d}{10}\right)^5\left(\frac{m_{\wt \nu_1}}{100\rm GeV}\right).
\end{align}
For a light ADM, generically we require quite small couplings $y_{i1}\lesssim10^{-2}$ except for much heavier neutralinos, says 10 TeV or above. This indicates that the left-handed sneutrino fraction in $\wt \nu_1$ should be highly suppressed. For heavy ADM around the weak scale, we need even smaller $y_{i1}\lesssim10^{-3}$ for reasonably heavy neutralinos.

To decouple CVS processes as early as possible, at least one of the three options should be relied on: 1) A quite small $y_{N}$;  2) quite heavy neutralinos; 3) ADM is dominated by $\wt{\nu}'_L$, namely $\cos\wt\theta\ll1$. 

To end up this part we would like to make a comment on the value of $x_d$. If $x_d$ is within the region $(10,x_f)$, the Boltzmann suppression factor in Eq.~(\ref{asymmetry}) will be too significant. The resulting ``initial" ADM asymmetry (with respect to the stage of ADM oscillation) becomes
\begin{align}\label{eta0}
\eta_{0}\equiv Y_+(x_d)-Y_-(x_d)=f_{\rm ADM}(x_d)\frac{B(T_{sph})}{S_{\rm ADM}(T_{sph})}=-\frac{18}{11k}f_{\rm ADM}(x_d)\eta_b,
\end{align}
with $Y_{\pm}=n_\pm(x_d)/s(x_d)$ the comoving number densities of DM and anti-DM, respectively. To derive the above relation we have used the assumption that asymmetry in the neutrino sector does not change from $T_{sph}$ down to $T_d$. Therefore, $x_d$ should take a value not far above 1.

\subsubsection{$T\sim T_f$: chemical equilibrium breaking and symmetrically annihilating}\label{singlet}

At this stage, the sneutrino DM symmetrically annihilate. In order to get rid of the symmetric part, the annihilation cross section 
should be at least a few pb~\cite{Graesser:2011wi}. Restricted to the model specified in Section~\ref{model}, the sneutrino ADM 
fails to having large enough annihilation rate. The reason is attributed to nothing but just the one discussed in the last part, i.e., 
in order to make the sneutrino as a viable ADM candidate,  the CVS processes have to be decoupled as early as possible, which 
in turn make the annihilation rate very low. Therefore, to save the scenario, new sizable couplings are necessary introduced
for the RHN-like sneutrinos. Here we consider an economical way by introducing a singlet $S$ which couples to RHNs via
\begin{align}\label{}
W_s=&\ld_s SNN^c+\frac{M_S}{2}S^2...,\cr
{\cal L}_{s}^{soft}=&m_S^2|S|^2+\left(\frac{1}{2}B_sM_sS^2+\ld_s A_sS\wt \nu_R^*\wt\nu_L'+c.c.\right)+...,
\end{align}
where dots collect other irrelevant terms involving $S$. Note that it is important to impose $R-$parity under which the scalar/fermionic component of $S$ is even/odd; this symmetry could forbid the coupling like $SLH_u$ which leads to sneutrino LSP decay and moreover modify the seesaw structure. It is tempting to identify $S$ as the one in the next-to minimal supersymmetric standard model~\cite{Kang:2011wb}, but we leave it for further investigation and here we focus on the looser situation where $S$ has free mass and merely the above sizable coupling. 

Depending on the mass spectrum and size of couplings, there are quite a few ways to enhance the annihilation rate of 
$\wt\nu_1$. For instance, consider the simplest case with a light CP-even singlet scalar from $S=(S_R+IS_I)/\sqrt{2}$. Note that 
the fermonic component of $S$ is R-parity  odd and thus $M_S>m_{\wt\nu_1}$. But one can still get a much lighter $S_R$ via a 
properly large $B_s-$term which splits the mass degeneracy between $S_R$ and $S_I$. Moreover, $S_R$ couples to $\wt\nu_1$ 
through the term
\begin{align}\label{}
-{\cal L}_{\wt\nu_1}\supset \mu_{R11}S_R|\wt\nu_1|^2,\quad \mu_{R11}=\frac{\ld_sA_s}{\sqrt{2}}\sin2\wt\theta.
\end{align}
This term enhances the annihilation cross section of $\wt\nu_1\wt\nu_1^*\rightarrow S_RS_R$ mediated by $\wt\nu_1$ in the $u/t-$channel, which mitigates the reliance on large $\ld_s$. This way works only for the well mixed $\wt\nu_L'$ and $\wt\nu_R$, otherwise one may have to fall back on the contact interaction $\frac{1}{2}\ld_s^2S_R^2|\wt\nu_1|^2$ or annihilating into a pair of RHNs via singlino exchanging. Their cross sections scale as $\sigma v\sim \ld^4_s/(64\pi m_{\wt \nu_1}^2)$ and thus both require $\ld_s\sim1$. In any case, ADM is disfavored to be near or even above the TeV scale, except that one can tolerate $\ld_s$ significantly larger than 1. Hereafter we will focus on the favored case with ADM around the weak scale. 

\subsection{Sneutrino-antisneutrino oscillating and symmetry regeneration}
The global $U(1)_L$ is not an exact symmetry for sneutrino ADM and its tiny breaking leads to an important consequence, tiny 
mixing between DM and antiDM fields~\footnote{Or the opposite CP-eigenstates which are treated as two independent and 
degenerate flavors.} and as well their mass splitting. But at the early universe DM are produced in the CP-eigenstates and thus 
the mixing and splitting renders two DM and anti-DM oscillating. This phenomena was first touched in~\cite{osci0}
 and then was systematically studied by several groups in a model independent way~\cite{oscillating,Tulin:2012re,oscillating1}. To our knowledge, the M$L$SIS provides the best example in the sense of theoretical motivations.

\subsubsection{$U(1)_L$-violation and sneutrino mass splitting}
The mass splitting between DM and anti-DM plays a center role for ADM oscillation. 
After the rotation specified by Eq.~(\ref{rotation}) and taking into account the operators that break $U(1)_L$, one has the following sneutrino mass terms
\begin{align}\label{}
{\cal L}\supset m_{\wt\nu_1}^2|\wt\nu_1|^2+m_{\wt\nu_2}^2|\wt\nu_2|^2+\left(\frac{1}{2}\delta m_{11}^2(\wt \nu_1^*)^2+\frac{1}{2}\delta m_{22}^2(\wt\nu_2^*)^2+\delta m_{12}^2\wt \nu_1^*\nu_2^*+c.c.\right)
\end{align}
with DM number or $U(1)_L$ violating mass parameters given by
\begin{align}\label{}
\delta m_{11}^2&\approx -m_NM_{N}\sin2\wt\theta-B_{M}M_N\sin^2\wt\theta,\cr
\delta m_{22}^2&\approx m_NM_{N}\sin2\wt\theta+B_{M}M_N\cos^2\wt\theta, \cr
 \delta m_{12}^2&= m_NM_{N}\cos2\wt\theta-\frac{1}{2}B_{M}M_N\sin^2\wt\theta.
\end{align}
In the decoupling limit $\wt\theta\rightarrow0$, the mass splitting among the CP-even and -odd components of $\wt\nu_1=\frac{1}{\sqrt{2}}\left({\rm Re}\wt\nu_1+I {\rm Im}\wt\nu_1\right)$ is 
\begin{align}\label{deltam}
\delta m\approx \frac{\delta m_{11}^2}{m_{\wt\nu_1}}= \frac{-m_NM_{N}\sin2\wt\theta-B_{M}M_N\sin^2\wt\theta}{m_{\wt\nu_1}}.
\end{align}
At leading order the splitting is independent of $\delta m_{12}^2$ and $\delta m_{22}^2$. $\wt\nu_L'$ furnishes the source of $U(1)_L$ violation and transfers it to other sneutrinos $\wt\nu_R$ and $\wt\nu_L$ through mixing. Therefore, if ADM $\wt\nu_1$ is dominated by $\wt\nu_R$, mass splitting between the components of $\wt\nu_1$ will be suppressed by small mixing.

Obviously, $\delta m$ can not be too large, otherwise the oscillation will happen too early and ruin the asymmetric DM scenario. Later we will specifically discuss how small $\delta m$ is required. Given a single family of $(N,N^c)$, the order of mass splitting typically should be not much below the active neutrino mass scale as long as the mixing angle is not extraordinarily small. This can be seen from Eq.~(\ref{deltam}), for the weak scale soft terms one has (assuming the first term can be made arbitrarily small) 
\begin{align}\label{}
\delta m\sim \left( B_M/m_{\wt\nu_1}\right) M_N \sin^2\wt\theta \sim \left( \frac{m_N}{m_D}\sin\wt\theta\right)^2 m_\nu,
\end{align}
where we have assumed that both $B_M$ and $m_{\wt \nu_1}$ are around the weak scale thus $B_M/m_{\wt\nu_1}\sim 1$. On the other hand, the largest neutrino mass scale is $\sim0.1$ eV, which means that the resulting $\delta m$ typically is a few orders larger than this scale except for a very small $\sin\wt\theta$. Soon later we will show that such an estimated $\delta m$ is many orders larger than the maximally tolerated mass splitting by sneutrino oscillation. So, we are led to conjecture that there is one splitting family of neutrino with mass hierarchically lighter than others and the sneutrino ADM candidate is dominant by the corresponding superpartner.

\subsubsection{Sneutrino oscillation}

To describe the evolution of densities with oscillation, 
 we follow the coupled Boltzman equations (BEs), treating DM and anti-DM as two coherent flavors with comoving number density matrix:  
\begin{align}\label{}
Y(x)=\left(\begin{array}{cc} Y_{11}(x) & Y_{12}(x) \\ Y_{21}(x) & Y_{22} (x)\\
 \end{array}\right),
\end{align}
where $``{11}"$ and $``22"$ denotes the DM and anti-DM flavor, respectively. For the diagonal elements one considers the quantities $Y_{\pm}=Y_{11}\pm  Y_{22}$ and for the coherent (off-diagonal) elements one consider $Y_{c\pm}=Y_{12}\pm Y_{21}$ with $Y_{c+}$ identical to zero in the absence of elastic scattering effects. The BEs for $Y_{ij}$ are first derived via the direct analogy to the neutrino oscillation~\cite{oscillating1} and 
then improved using the density matrix method~\cite{oscillating,Tulin:2012re}. In particular, the latter points out the irrelevance of elastic scattering if DM-plasma interactions are flavor-blind, as is just the case for the sneutrino ADM. Actually, in this case their results coincide and both give (with some small modifications here)
\begin{align}\label{}
Y'_+(x)=&-2\frac{\langle\sigma v \rangle s(x)}{x H(x)} \left[    \frac{1}{4} \left( Y^2_+(x)-Y^2_-(x)+Y_{c-}^2(x) \right)-Y^2_{eq}(x)  \right],\\
Y'_-(x)=&2\frac{\delta m}{x H(x)}Y_{c-}(x) , \label{BE2}\\
Y'_{c-}(x)=&-2\frac{\delta m}{x H(x)}Y_-(x)- \frac{\langle\sigma v \rangle s(x)}{x H(x)} Y_{c-}(x) Y_+(x),
\end{align}
where the Hubble expansion rate is rewritten as $H(x)={1.66 g_*^{1/2}m_{\rm DM}^2}x^{-2}/{M_{\rm Pl}}\equiv H_m/x^2$; the comoving number density in thermal equilibrium is
\begin{align}\label{}
Y_{eq} (x)= \frac{45}{2 \pi^4} \sqrt{\frac{\pi}{8}} \frac{g}{g_{*S}} x^{3/2} e^{-x}.
\end{align}
The eventual relic density of ADM is 
\begin{align}\label{}
\Omega_{\rm DM}h^2\approx 2.82\times 10^{10}Y_{+}(x\rightarrow\infty)\left(\frac{m_{\wt \nu_1}}{100\rm GeV}\right).
\end{align}

Several comments are in orders. First, the equation\eqref{BE2} among the BEs indicates that the asymmetry stays at its initial value $Y_-(x_0)\equiv \eta_0$ as long as $\delta m$ is negligible. Second, the thermally averaged annihilation cross section $\langle\sigma v \rangle$ is assumed to be a constant $\sigma_0$, with free value. Last, a rough estimation about the temperature at which oscillation commences is used~\cite{oscillating}:
\begin{align}\label{}
x_{osc}\approx \left( \frac{H_m\sigma_0 s_m \eta_0/2}{\delta m^2}\right)^{1/5}\sim 7.6\left(\frac{m_{\rm DM}}{400\rm GeV}\right) \left(\frac{10^{-5}\rm eV}{\delta m}\right)^{2/5}\left(\frac{g_{*S}}{10}\sqrt{\frac{g_*}{10}}\frac{\sigma_0}{10\rm pb}\frac{\eta_0}{0.1\eta_B} \right)^{1/5}.
\end{align}
We have chosen a smaller $\eta_0$ on account of the difficult in decoupling and hence Boltzman suppressed $\eta_0$ in Eq.~(\ref{eta0}). The above estimation shows that, in order to accommodate a larger $\delta m$, a smaller $x_{osc}$ and especially heavier DM are strongly  favored. 
\begin{figure}[htb]
\begin{center}
\includegraphics[width=3.0in,natwidth=10,natheight=15]{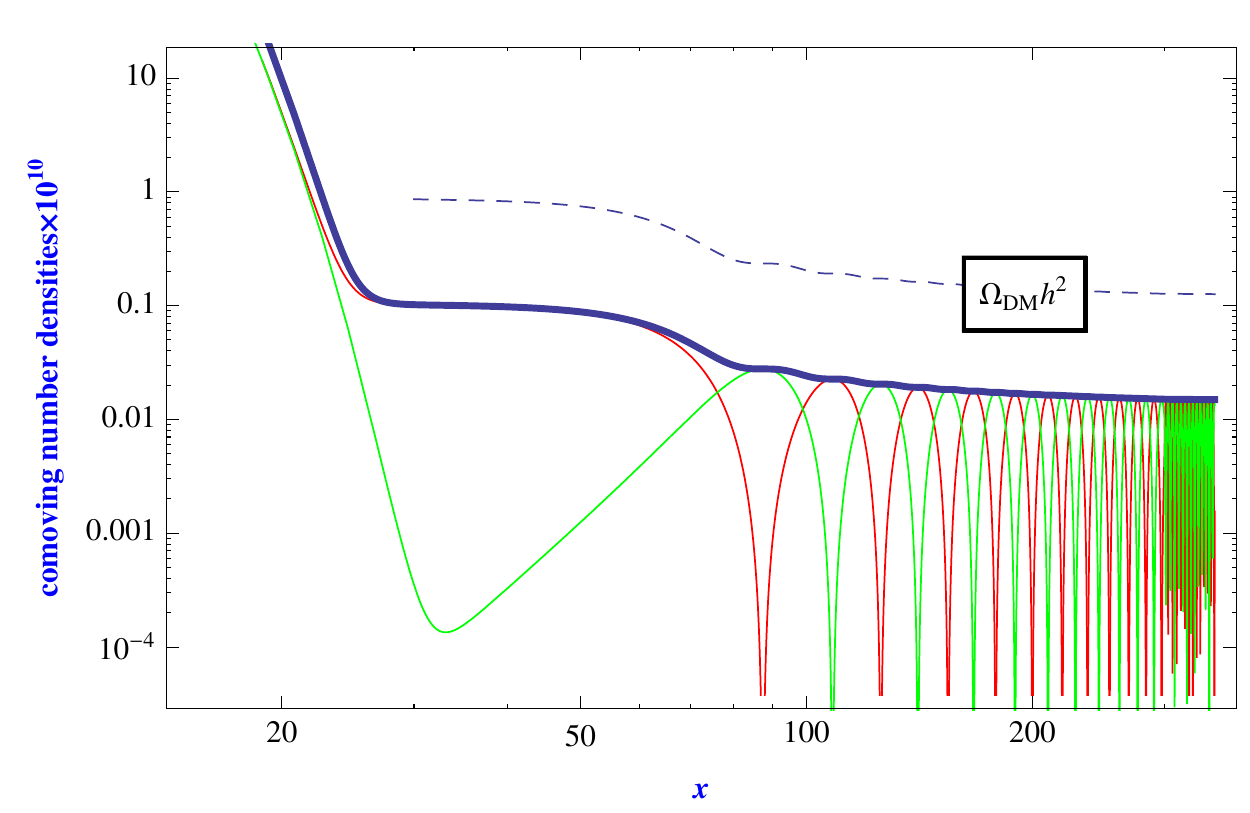}
\includegraphics[width=3.0in,natwidth=10,natheight=15]{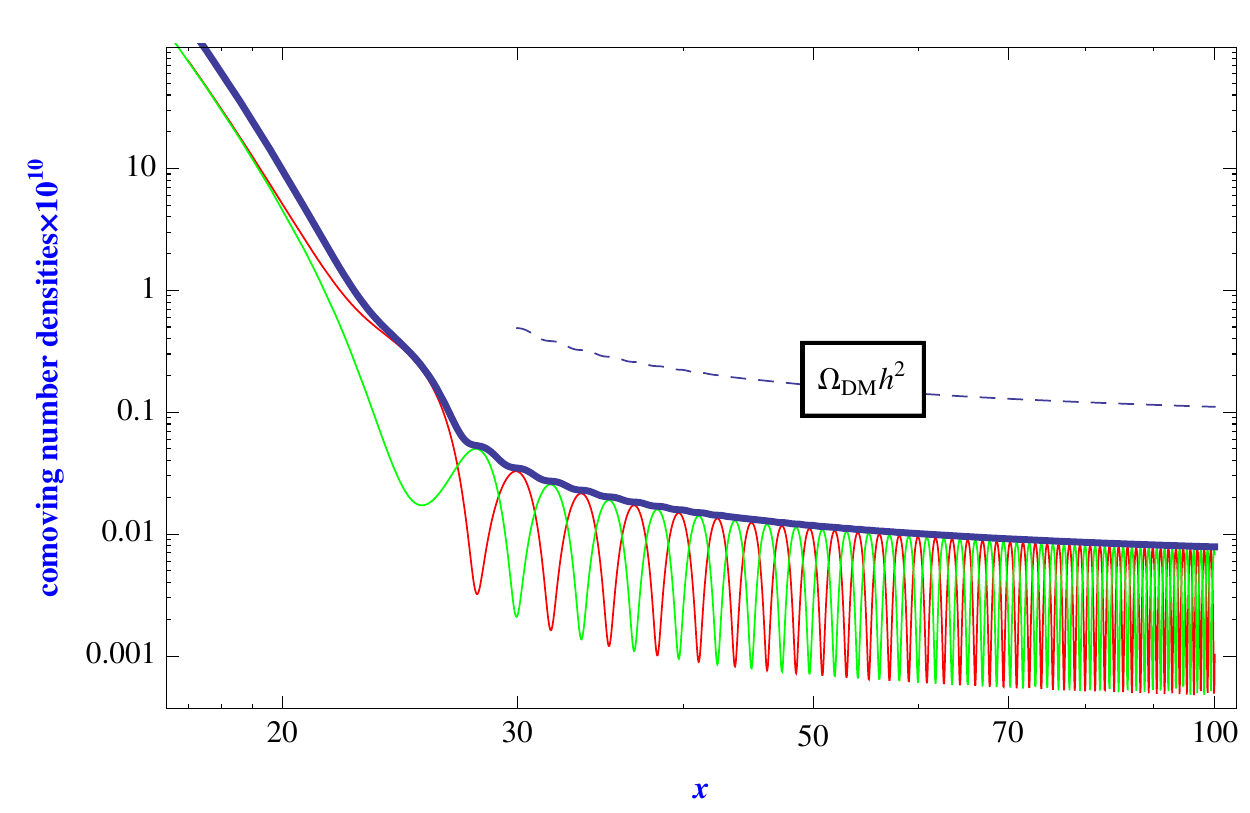}
\end{center}
\caption{Evolutions of the comoving densities of the quantities $Y_+$ (thick black line), $Y_{11}$ (red line) and $Y_{22}$ (green line). Parameters set for the Left: $\delta m=10^{-7}$ eV, $m_{\wt\nu_1}=300$ GeV, $\sigma_0=3$ pb, $\eta_0=0.1\eta_B$; Right: $\delta m=10^{-5}$ eV, $m_{\wt\nu_1}=500$ GeV, $\sigma_0=2$ pb, $\eta_0=0.5\eta_B$. In order to show the evolution of the total DM number density (dashed lines), we multiply the comoving densities by a factor $10^{10}$ in the vertical axis; eventually, the correct relic density $\Omega_{\rm DM}h^2\approx0.1$ is obtained for a sufficiently large $x$. 
 }
\label{fig2}
\end{figure}

We stress again that a heavy ADM produces a well consistent picture from several aspects. 
First, a larger $\delta m$ means that oscillation happens at earlier time, which helps to decrease the relic number density, resulting with a heavier ADM. At the other hand, a heavier sneutrino is good for suppressing $\delta m$, see Eq.~(\ref{deltam}). Finally, from the point view of model building as stated before, $\delta m\sim 10^{-10}$ eV is unappealing, while a value $10^{-5}$ eV, given $\sin\theta\lesssim10^{-2}$ along with a moderately small $B_M\sim {\cal O}({\rm GeV})\sim 10^{-2}m_{\wt\nu_1}$, is well 
acceptable.

We show the numerical results in Fig.~\ref{fig2}. We choose two set of benchmark points, (1) $\delta m=10^{-7}$ eV, $m_{\wt\nu_1}=300$ GeV, $\sigma_0=3$ pb, $\eta_0=0.1\eta_B$; (2) $\delta m=10^{-5}$ eV, $m_{\wt\nu_1}=500$ GeV, $\sigma_0=2$ pb, $\eta_0=0.5\eta_B$. In the left graph, the plateau is due to the ordinary freeze-out of ADM, but later the total DM density $Y_+$ decreases again as the oscillation commences. We can see that both cases generate correct DM relic density. 
Obviously, the allowed $\delta m$ is quite sensitive to the ADM mass. Doubling the ADM mass leads to two orders of magnitude increasing of the allowed range of $\delta m$.

\subsection{On the detections on sneutrino ADM}

As a remarkable difference than the ordinary ADM scenario, the oscillating ADM can generate indirect detection signatures. In Section~\ref{singlet} we have argued that the singlet $S$ allows a sufficient ADM annihilation rate via channels such as $\wt \nu_1\wt\nu_1^*\rightarrow S_R S_R$ (or annihilating into a pair of RHNs). The other channels are extremely suppressed owing to the CVS constraints; in particular the neutrino pair channel, which generates monochromatic neutrino signals, is still inaccessible even from the dwarf galaxies spiked by an intermediate massive black hole~\cite{Arina:2015zoa}. Therefore, the most likely signature from the sky is $\wt \nu_1\wt\nu_1^*\rightarrow S_R  S_R\rightarrow (b\bar b)(b\bar b) $, assuming that $S_R$ dominantly decays into a pair of bottom quarks through its mixing with the Higgs doublets (which are not explicitly given in the model because it is quite model dependent). It is shown that the Fermi dwarf limits provide the strongest constraint~\cite{Elor:2015bho}: for DM $\gtrsim 300$ GeV, the upper bound on the annihilation cross section of $\wt \nu_1\wt\nu_1^*\rightarrow S_R  S_R$ is at the pb level; on the other hand, we typically need a cross section of a few pb and thus the model is in the vicinity of exclusion\footnote{If $S_R$ dominantly decays into a pair of active neutrino, the bounds can be substantially relaxed.}.


As for the direct detection, we concentrate on the SM-like Higgs boson mediated DM-nucleon scattering (with cross section $\sigma_{p,h}=4m_p^2a_{p,h}^2/\pi $),  turning off the contribution from $S_R$ since its coupling to quarks in principle can be arbitrarily small. To estimate $\sigma_{p,h}$, we work in the decoupling limit of two Higgs doublets and as well $\tan\beta\gg1$, then having
\begin{align}\label{api}
a_{p,h}\approx 0.5\times10^{-3}\times\frac{\mu_{h 11}}{2m_{\wt\nu_1}}\frac{1}{m_{h}^2}~~ {\rm with }~~ \mu_{h11}=\sqrt{2}y_N^2\cos^2\wt\theta\, v_u,
\end{align}
with $v_u\approx174$ GeV. The bound Eq.~(\ref{CVSbound}) means typically $\mu_{h11}\lesssim10^{-4} v_u$ (taking $\sin\theta_\nu\sim 0.1$). As a result, we have $a_{p,h}\lesssim 10^{-12}\times \left(350{\rm GeV}/m_{\wt\nu_1}\right)\rm GeV^{-2}$ and therefore $\sigma_{p,h}\lesssim 10^{-15}$ pb, far below the current bounds from dark matter direct detection experiments. In addition, searching SUSY with sneutrino LSP is of particular interest at the LHC since it provides different signatures than those of the ordinary LSP scenario~\cite{Guo:2013asa, Arina:2015uea, Mitzka:2016lum}. In our scenario since the sneutrino ADM, as the LSP, is favored to be relatively heavy, we have a heavier SUSY spectra which still hides out.

\section{Conclusion}

Opposite to the canonical seesaw mechanism which introduces large lepton number $U(1)_L$ breaking by the heavy right-handed neutrinos, the inverse seesaw mechanism attributes the smallness of neutrino mass to a tiny $U(1)_L$ breaking, potentially allowing for a highly good $U(1)_L$. In this work we propose the maximally $U(1)_L$ supersymmetric inverse seesaw, in which unconventional dark matter phenomenologies arise when the singlet-like sneutrino is the lightest sparticle. It can be asymmetric DM due to the highly conserving of $U(1)_L$, but actually it is oscillating due to the slightly breaking of $U(1)_L$. To  maintain the ADM scenario, we find that the sneutrino is favored to be heavy near the weak scale instead of light around the GeV scale.

\section*{Acknowledgement}
The work is supported in part by the National Science Foundation of China under Grand No. 11175069 and 11422545.



\begin{thebibliography}{99}
\itemsep 0.5mm




\bibitem{Barr:1990ca}
  S.~M.~Barr, R.~S.~Chivukula and E.~Farhi,
  Phys.\ Lett.\  B {\bf 241}, 387 (1990);
  D.~B.~Kaplan,
  Phys.\ Rev.\ Lett.\  {\bf 68}, 741 (1992);
  N.~Cosme, L.~Lopez Honorez and M.~H.~G.~Tytgat,
  Phys.\ Rev.\  D {\bf 72}, 043505 (2005);
  R.~Kitano, H.~Murayama and M.~Ratz,
  Phys.\ Lett.\  B {\bf 669}, 145 (2008); 
  D.~E.~Kaplan, M.~A.~Luty and K.~M.~Zurek,
  Phys.\ Rev.\  D {\bf 79}, 115016 (2009);
  N.~Haba and S.~Matsumoto,
  arXiv:1008.2487 [hep-ph];
  H.~An, S.~L.~Chen, R.~N.~Mohapatra and Y.~Zhang,
  JHEP {\bf 1003}, 124 (2010); P.~H.~Gu, M.~Lindner, U.~Sarkar and X.~Zhang,
  Phys.\ Rev.\ D {\bf 83}, 055008 (2011); 
  Y.~Cui, L.~Randall and B.~Shuve,
  JHEP {\bf 1204}, 075 (2012);    Phys.\ Rev.\ D {\bf 84}, 123505 (2011);   H.~Davoudiasl, D.~E.~Morrissey, K.~Sigurdson and S.~Tulin,
  Phys.\ Rev.\ D {\bf 84}, 096008 (2011); 
  N.~F.~Bell, K.~Petraki, I.~M.~Shoemaker and R.~R.~Volkas,
  Phys.\ Rev.\ D {\bf 84}, 123505 (2011); 
  S.~M.~Barr,
  Phys.\ Rev.\ D {\bf 85}, 013001 (2012); 
  C.~Arina and N.~Sahu,  Nucl.\ Phys.\ B {\bf 854}, 666 (2012); 
K.~Petraki, M.~Trodden and R.~R.~Volkas,
  JCAP {\bf 1202}, 044 (2012); 
  W.~Z.~Feng, P.~Nath and G.~Peim,
  Phys.\ Rev.\ D {\bf 85}, 115016 (2012); 
  P.~H.~Gu,
  Nucl.\ Phys.\ B {\bf 872}, 38 (2013);  
  R.~T.~D'Agnolo and A.~Hook,
  Phys.\ Rev.\ D {\bf 91}, no. 11, 115020 (2015). 
A more complete reference list see the reviews~\cite{Petraki:2013wwa,Zurek:2013wia}. 
 
 \bibitem{Petraki:2013wwa} 
  K.~Petraki and R.~R.~Volkas,
  Int.\ J.\ Mod.\ Phys.\ A {\bf 28}, 1330028 (2013). 

\bibitem{Zurek:2013wia} 
  K.~M.~Zurek,
  Phys.\ Rept.\  {\bf 537}, 91 (2014)
  [arXiv:1308.0338 [hep-ph]].
 

\bibitem{Buckley:2010ui} 
  M.~R.~Buckley and L.~Randall,
  JHEP {\bf 1109}, 009 (2011)
  doi:10.1007/JHEP09(2011)009
  [arXiv:1009.0270 [hep-ph]].


\bibitem{Kang:2011ny} 
  Z.~Kang and T.~Li,
  JHEP {\bf 1210}, 150 (2012). 


\bibitem{Boucenna:2015haa} 
  S.~M.~Boucenna, M.~B.~Krauss and E.~Nardi,
  Phys.\ Lett.\ B {\bf 748}, 191 (2015).



\bibitem{Unwin:2012rp} 
  J.~Unwin,
  JHEP {\bf 1306}, 090 (2013)
  [arXiv:1212.1425 [hep-ph]].

\bibitem{Barr:2015lya} 
  S.~M.~Barr and R.~J.~Scherrer,
  arXiv:1508.07469 [hep-ph].



  \bibitem{Mohapatra:1986bd}
  R.~N.~Mohapatra and J.~W.~F.~Valle,
  Phys.\ Rev.\  D {\bf 34}, 1642 (1986);
  M.~C.~Gonzalez-Garcia and J.~W.~F.~Valle,
  Phys.\ Lett.\  B {\bf 216}, 360 (1989).


  
  \bibitem{Kang:2011wb} 
  Z.~Kang, J.~Li, T.~Li, T.~Liu and J.~Yang,
  arXiv:1102.5644 [hep-ph].

\bibitem{Mitropoulos:2013fla} 
  P.~Mitropoulos,
  JCAP {\bf 1311}, 008 (2013)
  doi:10.1088/1475-7516/2013/11/008
  [arXiv:1307.2823 [hep-ph]].


\bibitem{Arina:2008bb}
  C.~Arina {\it et al.},
  Phys.\ Rev.\ Lett.\  {\bf 101}, 161802 (2008).
  
  \bibitem{An:2011uq} 
  H.~An, P.~S.~B.~Dev, Y.~Cai and R.~N.~Mohapatra,
  Phys.\ Rev.\ Lett.\  {\bf 108}, 081806 (2012)
  [arXiv:1110.1366 [hep-ph]].


\bibitem{Guo:2013sna} 
  J.~Guo, Z.~Kang, T.~Li and Y.~Liu,
  JHEP {\bf 1402}, 080 (2014)
  [arXiv:1311.3497 [hep-ph]].
 
 
 \bibitem{Choi:2013fva} 
  K.~Y.~Choi and O.~Seto,
  Phys.\ Rev.\ D {\bf 88}, no. 3, 035005 (2013);   
  K.~Y.~Choi, O.~Seto and C.~S.~Shin,
  JHEP {\bf 1409}, 068 (2014). 


  
\bibitem{Hooper:2004dc}
  D.~Hooper, J.~March-Russell and S.~M.~West,
  Phys.\ Lett.\  B {\bf 605}, 228 (2005).

\bibitem{Page:2013fva} 
 V. Page, JHEP 0704 (2007) 021 [hep-ph/0701266].

\bibitem{oscillating} 
  M.~Cirelli, P.~Panci, G.~Servant and G.~Zaharijas,
  JCAP {\bf 1203}, 015 (2012). 


\bibitem{Tulin:2012re} 
  S.~Tulin, H.~B.~Yu and K.~M.~Zurek,
  JCAP {\bf 1205}, 013 (2012)
  [arXiv:1202.0283 [hep-ph]].

\bibitem{oscillating1} 
  M.~R.~Buckley and S.~Profumo,
  Phys.\ Rev.\ Lett.\  {\bf 108}, 011301 (2012). 
  
\bibitem{osci0}
T. Cohen and K. M. Zurek, Phys. Rev. Lett. 104 (2010) 101301; 
E. J. Chun, Phys. Rev. D 83 (2011) 053004; 
A. Falkowski, J. T. Ruderman, T. Volansky, JHEP 1105 (2011) 106; 
C. Arina, N. Sahu, [arXiv:1108.3967 [hep-ph]].


\bibitem{Antusch:2014woa} 
  S.~Antusch and O.~Fischer,
  JHEP {\bf 1410}, 94 (2014). 



\bibitem{TEU}
 V. Mukhanov, ``Physical Foundations of Cosmology".


\bibitem{baryon:as}
E. Komatsu et al. [ WMAP Collabo- ration ], Astrophys. J. Suppl. 192 (2011) 18.


\bibitem{Ellwanger:2012yg} 
  U.~Ellwanger and P.~Mitropoulos,
  JCAP {\bf 1207}, 024 (2012)
  [arXiv:1205.0673 [hep-ph]].
  

\bibitem{Graesser:2011wi} 
  M.~L.~Graesser, I.~M.~Shoemaker and L.~Vecchi,
  JHEP {\bf 1110}, 110 (2011). 
  
  
\bibitem{Arina:2015zoa} 
  C.~Arina, S.~Kulkarni and J.~Silk,
  Phys.\ Rev.\ D {\bf 92}, no. 8, 083519 (2015).

\bibitem{Elor:2015bho} 
  G.~Elor, N.~L.~Rodd, T.~R.~Slatyer and W.~Xue,
  JCAP {\bf 1606}, no. 06, 024 (2016).


\bibitem{Guo:2013asa} 
  J.~Guo, Z.~Kang, J.~Li, T.~Li and Y.~Liu,
  JHEP {\bf 1410}, 164 (2014).
  
  \bibitem{Arina:2015uea} 
  C.~Arina, M.~E.~C.~Catalan, S.~Kraml, S.~Kulkarni and U.~Laa,
  JHEP {\bf 1505}, 142 (2015).

  
  \bibitem{Mitzka:2016lum} 
  L.~Mitzka and W.~Porod,
  arXiv:1603.06130 [hep-ph].









 
 
  
    
  
  
  


 




 



\end{thebibliography}
\end{document}